\documentstyle[12pt,aasms4]{article}
\slugcomment{Submitted to The Astrophysical Journal}
\lefthead{Nemiroff}
\righthead{Bright Lenses and Optical Depth}

\begin{document}
\title{Bright Lenses and Optical Depth}
\author{Robert J. Nemiroff}
\affil{Department of Physics, Michigan Technological University,
       Houghton, MI  49931}

\begin{abstract}
In gravitational lensing, the concept of optical depth assumes the lens
is dark.  Several microlensing detections have now been made where the
lens may be bright.  Relations are developed between apparent and
absolute optical depth in the regime of the apparent and absolute
brightness of the lens.  An apparent optical depth through bright
lenses is always less than the true, absolute optical depth.  The
greater the intrinsic brightness of the lens, the more likely it will
be found nearer the source.  \end{abstract}


\section{Introduction}

Optical depth to gravitational lensing is a measurement of lens surface
density (Vietri \& Ostriker 1983).  More specifically, the optical
depth $\tau$ to Schwarzschild lenses is the angular fraction of the sky
inside their Einstein radii.  The angular Einstein radius has magnitude
$\theta_E = \sqrt{2 R_S D_{LS} / D_{OL} D_{OS}}$ where $R_S$ is the
Schwarzschild radius of the lens, $D$ is angular diameter distance, and
subscripts $O$, $L$, and $S$ refer to the observer, lens and source,
respectively (see, for example, Refsdal 1964).  When the lens is inside
an Einstein ring centered on the source, the resulting magnification is
greater than $\sim$1.34.

In the past few years, hundreds of microlensing detections have been
claimed by several collaborations actively seeking out such events
(Paczynski et al. 1986; Alcock et al. 1993; Aubourg et al. 1993;
Udalski et al. 1993). These detections have been converted into
estimates of the optical depth to lensing through our Galactic halo and
disk (Bennett et al. 1995; Gates, Gyuk, and Turner 1995).  Currently
much debate exists about the optical depth of our Galactic halo, as
well as the implied masses for the lens events (Alcock et al. 1996;
Jetzer 1994; Han and Gould 1996; Gould 1996) and the implied mass
density of the lenses.

The addition of any unlensed component to a source brightness
determination is called ``blending" (Griest and Hu 1992; Kamionkowski
1995; Di Stefano \& Esin 1995; Buchalter and Kamionkowski 1997).  For
lenses below about 10$^6$ $M_{\odot}$, source and lens images are too
close to resolve by normal telescopic means (Gould 1992).  There are,
however, several ways to detect the presence of an unlensed component.
Photometrically, a blended light curve has a different shape than a
unblended one, even for a single lens (Pratt 1994).  Therefore, at
least theoretically, the brightness of the unlensed component can be
deconvolved (Buchalter, Kamionkowski, and Rich 1996).  In practice,
however, the photometry needs to be increasingly good to detect an
increasingly faint unlensed component (Wozniak and Paczynski 1997).

At least two other methods of detecting an unlensed component exist.
An unlensed component is unlikely to have an identical spectrum as the
lensed component, so lens amplification will change the relative
contribution to the total light: a color change (Kamionkowski 1995).
Also, were the lens a binary creating at any time 5 or more source
images, a drop below a given minimum brightness would unambiguously
indicate the presence of an unlensed component (Witt and Mao 1994).  In
this paper it is assumed that the amount of blending could not be
determined from any of the above methods.

The information content in any unblended microlensing event is
contained in only three variables: the maximum amplification $A_{max}$,
the duration $t_{dur}$, and the time of maximum light $t_{max}$
(Nemiroff \& Wickramasinghe 1994).  In theory, optical depth can be
computed solely from the distribution of these three parameters -
including no information about unlensed components.  In practice, an
optical depth reported may include a model for source blending (Pratt
et al. 1994, Alcock et al. 1995). Such a model may include significant
assumptions about the luminosity function of potentially blended
sources and specifics of realistic observing programs (Buchalter,
Kamionkowski, and Rich 1996).  Typically, no model is included for the
luminosity function of the potentially blended lenses, however.

This paper studies optical depth and differential optical depth in the
regime of the significantly bright lens.  Previous work is extended to
specifically include relations between the optical depth for dark
lenses and the optical depth for bright lenses.  The generated
relations are also relevant to specific cases of a dark lens and
multiple, unresolved, blended, sources.  Although exploring a purely
theoretical limit, practical application to present microlensing claims
will be briefly touched on in Section 3.

\section{Apparent Lens Luminosity and Optical Depth}

Were a gravitational lens to have apparent luminosity $l_L$ equal to an
unlensed source luminosity $l_S$, a measured 'apparent' source light
amplification $A_{app}$ of 1.34 would only result from a true or
'absolute' amplification $A_{abs}$ of 1.62.  This example shows that
the brightness of the lens dilutes the true apparent magnification.
More generally,
  \begin{equation} 
         A_{abs} = A_{app} (1 + l_L/l_S) - l_L/l_S .
  \end{equation} 
Figure 1 gives a plot of $A_{abs}/A_{app}$ versus
$l_L/l_S$.  Note that for small $l_L/l_S$ (dark lenses), the apparent
amplification $A_{app}$ is a good measure of absolute amplification
$A_{abs}$.  For apparently very bright lenses $(l_L >> l_S)$, however,
absolute amplifications may be much greater than the apparent
amplifications, by as much as the factor $l_L/l_S$.  This relation
holds even when subscript $L$ is interpreted as labeling an unresolved
source component.

When the apparent lens amplification differs from the absolute lens
amplification, the apparent optical depth $\tau_{app}$ differs from the
absolute optical depth $\tau_{abs}$.  Here {\it absolute} optical depth
$\tau_{abs}$ corresponds to the probability of {\it absolute} lens
amplification above 1.34, whereas {\it apparent} optical depth
$\tau_{app}$ corresponds to the probability of {\it apparent} lens
amplification above 1.34.  To compute $\tau_{abs}$ from the measured
$\tau_{app}$ and $l_L/l_S$, one can consider the case where $A_{app}$
is 1.34.  Given $A_{app}$ and a lens distance $D_L$, the lens must be
inside some radius $b < b_{app}$ to generate $A > A_{app}$.  The
relation between $b$ and $A$ is given in Nemiroff (1989):
  \begin{equation} 
         b = \sqrt{4 R_S D_{OL} D_{LS} \Phi / D_{OS}} ,
  \end{equation} 
where $\Phi = \sqrt{ A^2 / (A^2 - 1)} - 1$.  Now
Figure 1 shows that for any $l_L/l_S$, $A_{abs} > A_{app}$, so a more
perfect lens and source alignment must exist to create the higher true
$A_{abs}$.  (In other words, the bright lens must working harder to be
seen over its own brightness.)  So the corresponding radius $b_{abs}$
can be computed from equation (2), which  generates $A_{abs}=1.34$, is
necessarily less than $b_{app}$.  For any lens and source distance, $b
\propto \sqrt{\Phi}$, and since $\tau \propto b^2$, $\tau \propto
\Phi(A)$.  Therefore,
  \begin{equation} 
         \tau_{abs} = \tau_{app} \Phi_{abs} / \Phi_{app} .
  \end{equation} 
For large $A$, $\Phi \sim 1/(2A)$ so $\tau_{abs} \sim
\tau_{app} A_{app}/A_{abs}$.  For large $A$ and a very bright lens
($l_L >> l_S$), $A_{abs} / A_{app} \sim l_L/l_S$ from equation (1), so
then $\tau_{abs} \sim \tau_{abs} l_L / l_S$.

The line marked ``Apparent" in Figure 2 shows the relation between
$\tau_{abs}/\tau_{app}$ and apparent relative luminosity of the lens:
$l_L/l_S$.  As with the relationship between $A_{abs}/A_{app}$ and
$l_L/l_S$, for small $l_L/l_S$ (dark lenses), the apparent optical
depth $\tau_{app}$ is a good measure of absolute optical depth
$\tau_{abs}$.  For apparently very bright lenses $l_L >> l_S$, however,
absolute optical depth may be much greater than the apparent optical
depth.

\section{Absolute Lens Luminosity and Optical Depth}

How is optical depth affected by a field of bright lenses all with the
same {\it absolute} luminosity $L_L$?  Given both $L_L$ and the
absolute source luminosity $L_S$, one must also know the relative lens
and source distances to determine the relationship between $A_{abs}$
and $A_{app}$.  Given equation (1) and the $L \propto l/D^2$, it is
clear that 
 \begin{equation}
    A_{abs} = A_{app} \left( 1 + { L_L D_{OS}^2  \over L_S D_{OL}^2 }
    \right) - { L_L D_{OS}^2  \over L_S D_{OL}^2 } .
 \end{equation} 
Now a dark lens must fall into an ellipsoidal detection
volume to create an absolute amplification of a given source by an
amount $A_{abs}$ (Nemiroff 1989, Griest 1991, Nemiroff 1991).  But what
volume must a bright lens fall into to create an apparent amplification
of an amount $A_{app}$?  At each relative lens and source distance,
equations (2) and (4) were solved computationally to generate three
such canonical volumes, which are shown in Figure 3.  Here, the plotted
cross-sectional boundaries surround the volume a lens of intrinsic
brightness $L_L/L_S$ = 1/100, 1, and 100 must fall into to create an
apparent amplification greater than 1.34.

Inspection of Figure 3 leads to several conclusions.  First, it
indicates that $V$ is a monotonically decreasing function of
$L_L/L_S$.  In other words, the brighter the lens, the smaller the
detection volume, the ``harder" it is to detect the lens over its own
brightness.

Next, the shape of each detection volume shows that, in general,
$dV/dD_{OL}$ has a maximum nearer the source than the observer.
Therefore, although the bright lens could be at any distance between
the observer and the source and still be detected, the single most
detectable bright lens placement is closer to the source than the lens.
This is in direct contrast to dark lenses, which are most likely to be
detected precisely midway between the observer and the source.

Additionally, most of the space in the detection volume is nearer the
source.  Therefore, for uniformly distributed lenses, brighter lenses
are more likely to be detected closer to the source than the observer.
This is again in contrast to dark lenses, which have symmetric
detection probability about the midway point between the lens and the
source.

The relative change in the shapes of the detection volumes as absolute
lens brightness changes indicates that the maximum of $d/dL_L
(dV/dD_{OL})$ increases with $D_{OL}$.  In other words, for any spatial
distribution of lenses, increasingly (absolute) brighter lenses are
increasingly more likely to be detected nearer the source.

For a lens of a given absolute luminosity, it is, or course, more
probable to detect a low amplification event than a high amplification
event.  This is shown graphically in Figure 4 for $A_{app} = 0.1$,
1.34, and 10.  To generate this plot, equations (2) and (4) were again
solved computationally, but this time for constant $L_L/L_S$ (instead
of for $A_{app}=1.34$). Inspection of this plot shows, first, that $V$
is a decreasing function of $A_{app}$, so that increasingly higher
amplitudes yield increasingly smaller detection volumes and are
therefore increasingly less likely to be detected.  Although the shape
of each $A_{app}$ detection volume is not exactly a scaled version of
other $A_{app}$ detection volumes, the difference is only slight.
Therefore, although high $A_{app}$ events have a slightly larger
fraction of their volume at higher $D_{OL}$, the relative likely
placement between the observer and the source of a lens causing a high
$A_{app}$ is approximately the same as with a low $A_{app}$ event.

It should be noted that the actual probability of detecting a lens at
any distance from the observer is directly proportional to
$n_L(D_{OL})$, the density of lenses at that distance.  However, for
any lens distribution, relatively lower $A_{app}$ events will be found,
on average, shifted relatively closer to the source.

The relationship between $\tau_{abs}$ and $\tau_{app}$ for a field of
intrinsically bright lenses is defined similarly as for a field of
apparently bright lenses.  Given a relative lens and source distance,
one again must find the relative radii $b_{app}$ and $b_{abs}$ the lens
be placed from the observer-source axis to create the canonical
$A_{app}=1.34$ and $A_{abs}=1.34$ magnification that defines optical
depth.  From integrating $\pi b^2$ for all lens positions near the
observer - source axis, one finds the volume of the respective
detection volumes.  If the lens density is constant between the
observer and the source, these volumes are directly proportional to
$\tau$ (Nemiroff 1989).

The line marked ``Absolute" in Figure 2 shows the relation between
$\tau_{abs}/\tau_{app}$ and absolute relative luminosity of the lens:
$L_L/L_S$.  As expected, for small $L_L/L_S$ (dark lenses), the
apparent optical depth $\tau_{app}$ is a good measure of absolute
optical depth $\tau_{abs}$.  For absolutely very bright lenses $L_L >>
L_S$, however, absolute optical depth may be much greater than the
apparent optical depth.  Because of the divergence between $A_{abs}$
and $A_{app}$ at small $D_{OL}$, the discrepancy between $\tau_{abs}$
and $\tau_{app}$ is greater at a given $L_L/L_S$ than at the same level
of $l_L/l_S$.

\section{Practical Applications}

Although this paper is geared toward a better and more general
theoretical understanding of bright lenses, some of the above results
may be applied to specific recent claims that our Galaxy is composed of
a significant fraction of dark lenses. Recently reported results from
the MACHO and EROS microlensing collaborations indicate that much of
our Galactic halo is composed of lenses with mass on the order of a
fraction of a solar mass (Jetzer 1994; Alcock et al. 1996; Aubourg et
al. 1996).  Main sequence stars of this mass in our Galaxy would
normally be bright, and there is some debate as to whether they would
be detectable in HST surveys (Flynn, Gould, and Bahcall 1996).  Claims
that might be considered by some to be truly extraordinary have become
a standard interpretation of these results: that these lenses are {\it
dark}, implying a whole new class of astronomical objects.  Were these
lenses bright, then perhaps recent microlensing results could be
interpreted as more conventional objects.

Additionally, most microlensing detections are being made toward the
Galactic bulge (Udalski et al. 1993, Alcock et al. 1996).  In this
regime, a common lens is indeed a bright star, and hence the optical
depths in this direction too - which are significantly higher than
originally expected (Paczynski et al. 1994) -- might be affected by
better understanding of lens brightness.

This study might also give insight on how bright lenses could be and
still fall within  measured statistical determinations of measured
$A_{max}$, $t_{dur}$ and $t_o$.  From inspection of Figure 2, it is
evident that were the apparent brightness of the lenses 1/10th of the
source, the true optical depth would be a factor of 1.1 higher than the
published estimates.  This plot also holds for the case of dark lenses
and bright unresolved (blended) sources labeled ``L." Were the absolute
luminosity of the lenses 1/10th of the source, the true optical depth
would be a factor of about 1.6 higher than the published estimates.

One could turn this argument around and show that a limit on $l_L/l_S$
by other means (light curve shape, for example) would more clearly
indicate that a new class of dark lens must be invoked to explain the
optical depth inferred in our Galactic halo, even without considering
separate searches for the lenses (by HST, for example).

Alternatively, the above results could be used to bolster a more
conventional interpretation and distance to the lenses.  Were the
microlenses actually at the upper limit of possible brightness, the
above results indicate that it is more likely at least some of the
lenses are nearer the source than previously thought.  Given any lens
brightness, the likelihood is increased that the microlensing events of
source stars in the LMC are caused by bright lenses in the LMC itself,
and that stars in the Galactic bulge are caused by bright lenses in the
bulge itself.

Lastly, as blending is modeled in the optical depth fits of at least
one of the microlensing search groups (Alcock et al. 1995), there is
the possibility that systematic corrections might apply to account for
unknown attributes of the luminosity function of the lenses (or blended
sources) in published models.  It is hoped that this work gives insight
on the possible magnitude of such corrections.

\section{Summary and Conclusions}

Estimates of optical depth that do not include lens brightness always
underestimate the true optical depth.  For a field of identical objects
which act as both the lenses and sources, for example, the measured
optical depth will be only about 1/6th of the true optical depth.
Given that practically all optical depth estimates assume a dark lens,
true optical depth must be at least slightly higher than the published
estimates.

Not only does the optical depth change with lens brightness, but the
differential optical depth changes as well.  The resulting detection
volume is not symmetric between the observer and the source and has
more space nearer the source.  One consequence is that besides being
increasingly difficult to detect intrinsically brighter lenses, it is
increasingly difficult to detect them nearer the observer.  In other
words, the brighter the lens, the more likely it will be detected
nearer the source.

\clearpage

\figcaption[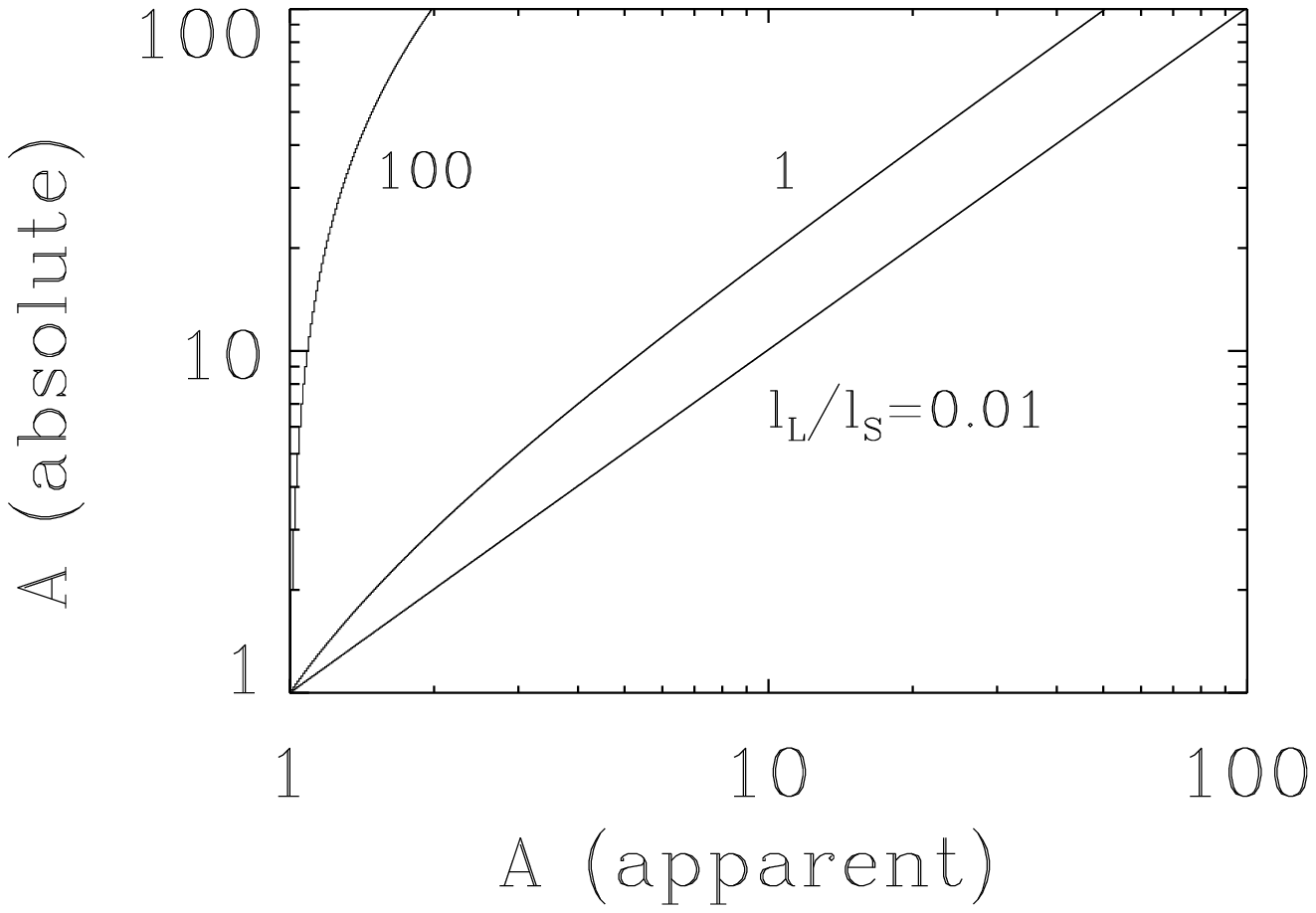]{A plot of absolute versus apparent amplification
for three values of the relative apparent brightness of the lens.
\label{fig1}}

\figcaption[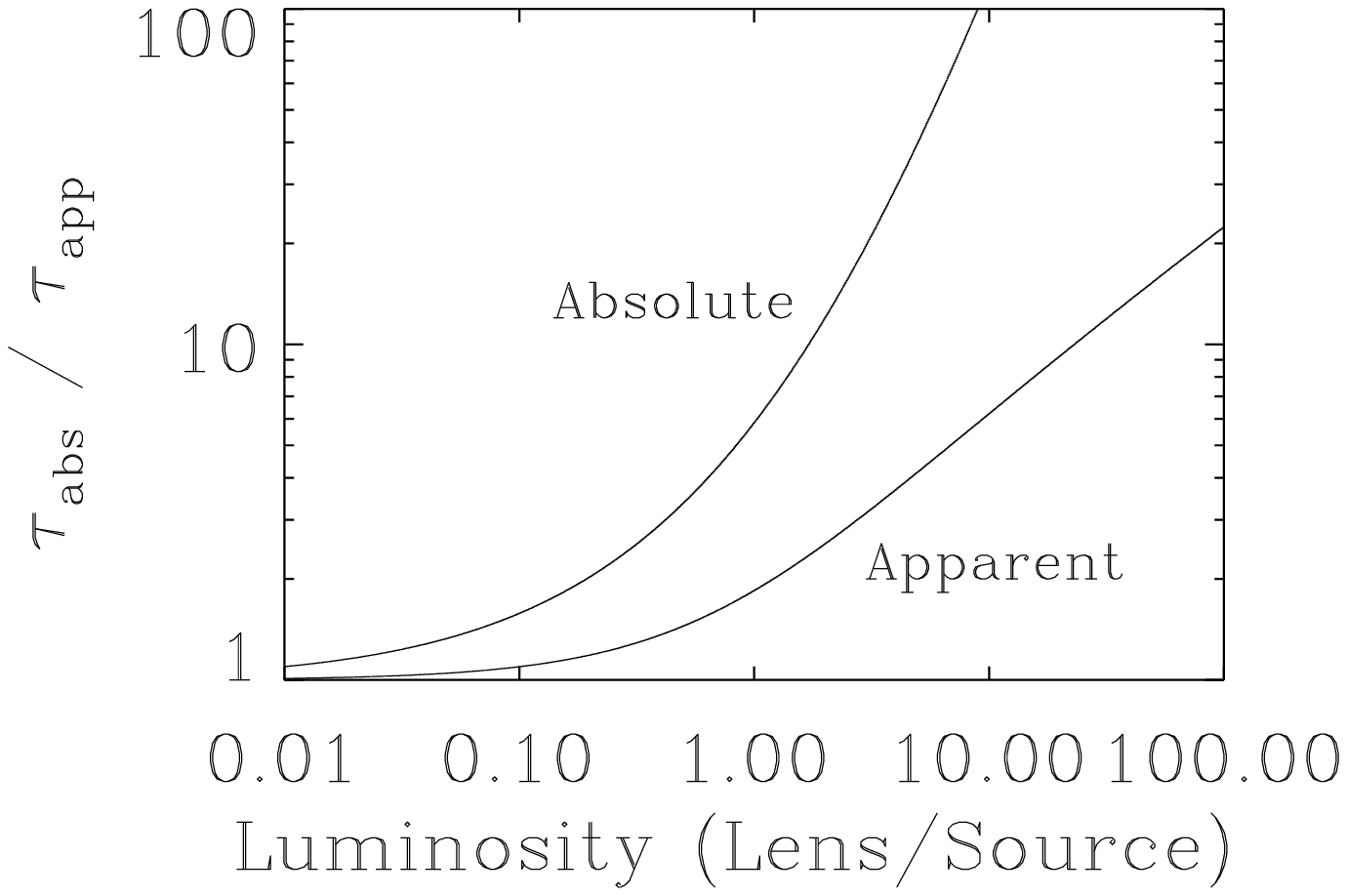]{Plots of `absolute' optical depth over
`apparent' optical depth for different levels of both apparent and
absolute brightness of the lens, relative to the source.  Apparent
optical depth is derived assuming the lens is dark.  In actuality, a
bright lens demands a higher `absolute' or true optical depth to
exist.  \label{fig2}}

\figcaption[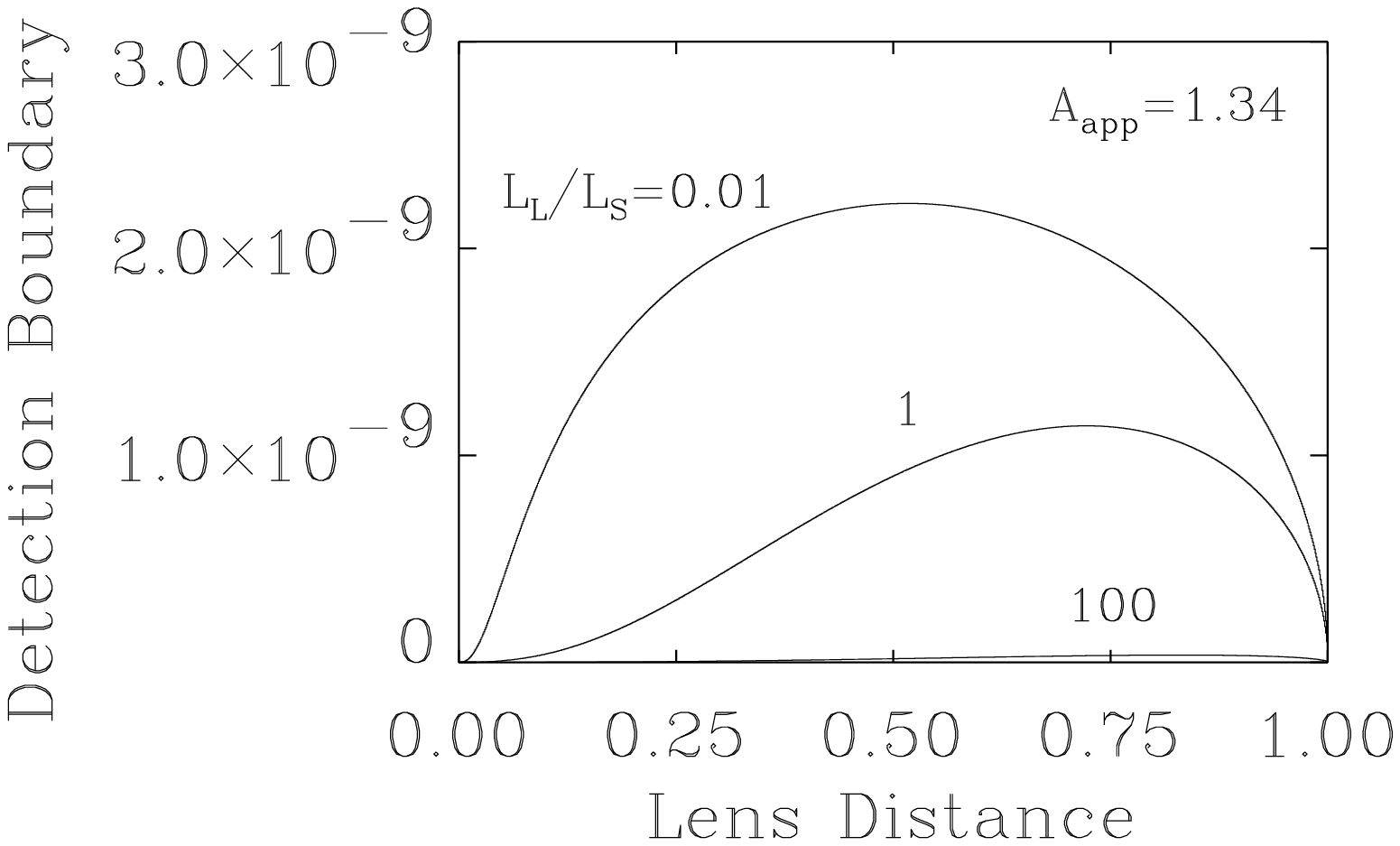]{Detection volume boundaries for different levels
of the absolute luminosity of the lens, relative to the absolute
luminosity of the source.  For a bright lens to be detected, it must
fall inside the detection volume boundary.  Note that the shape of the
detection volume becomes more asymmetric as lens brightness increases,
meaning that bright lenses are more likely detected nearer the source.
\label{fig3}}

\figcaption[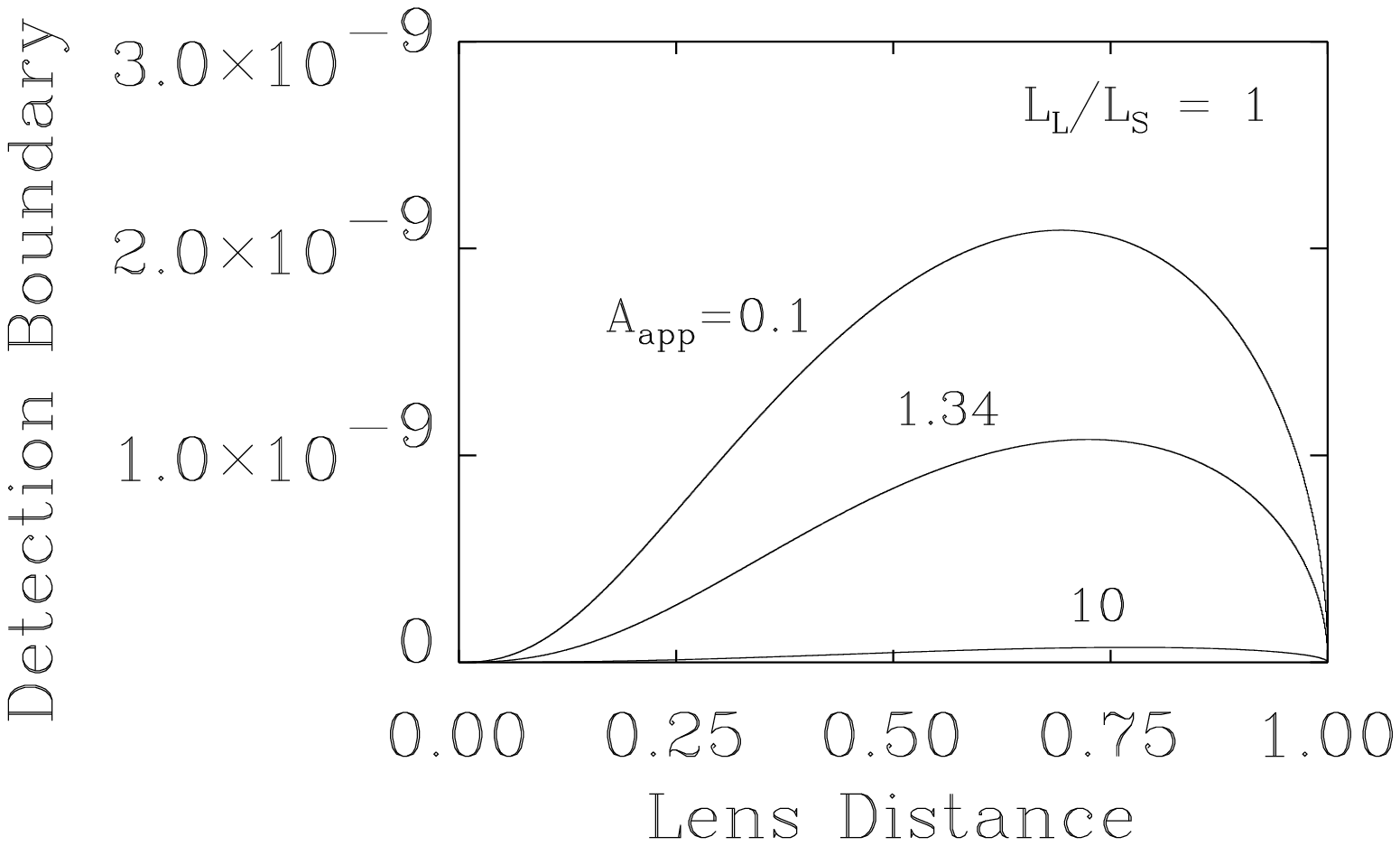]{Detection volume boundaries for different
amplifications, given a lens and source of equal absolute luminosity.
\label{fig4}}

\clearpage

\plotone{brtf1.eps}

\clearpage

\plotone{brtf2.eps}

\clearpage

\plotone{brtf3.eps}

\clearpage

\plotone{brtf4.eps}

\end{document}